\documentclass[5p]{elsarticle}
\usepackage{lineno,hyperref}
\modulolinenumbers[5]
\usepackage{graphicx}   
\usepackage{latexsym}   
\usepackage{bm}

\begin{document}

\begin{frontmatter}

\title{A question mark on the equivalence of Einstein and Jordan frames}

\author{Narayan Banerjee$^{1}$, Barun Majumder$^{2,3}$}

\address{$^1$Department of Physical Sciences, \\Indian Institute of Science Education and Research - Kolkata, \\West Bengal 741246, India.} 
\address{$^2$Department of Physics, Montana State University, Bozeman, MT 59717, USA} 
\address{$^3$Department of Physics, IIT Gandhinagar, Ahmedabad, India}

\date{\today}

\begin{abstract}
With an explicit example, we show that Jordan frame and the conformally transformed Einstein frames clearly lead to different physics for a non-minimally coupled theory of gravity, namely Brans-Dicke theory, at least at the quantum level. The example taken up is the spatially flat Friedmann cosmology in Brans-Dicke theory.
\end{abstract}


\end{frontmatter}



\section{Introduction}
Brans-Dicke theory\cite{brans} remained amongst the most talked about relativistic theory of gravity after general relativity. Soon after the theory was brought into being, a conformal transformation was suggested by Dicke\cite{dicke} which can recast the theory in a different frame, called Einstein frame where the field equations look more tractable. The original theory, given in a frame, popularly dubbed as the Jordan frame, is a manifestly non-minimally coupled theory where a long range scalar field $\phi$ has an interference term with the curvature $R$. The term looks like $\phi R$ in the action. A suitable conformal transformation of the form $\bar{g}_{\mu\nu} = \phi g_{\mu\nu}$ breaks this non-minimal coupling so that the scalar field contributes only through the kinetic term in the action. The matter part also now depends on the scalar field as the stress energy tensor goes through a corresponding transformation\cite{dicke}. This dependence does not formally affect the
  calculations unless one goes on to solve the geodesic equation. This conformally transformed frame is called the Einstein frame. Brans-Dicke (BD) theory indicates that the Newtonian constant $G$ effectively is a function of the scalar field $\phi$ as $G = \frac{G_{0}}{\phi}$, where $G_{0}$ is a constant and can be taken to be the present value of $G$ in a cosmological context. In the so called Einstein frame, $G$ regains its status of being a universal constant but the price one has to pay is that the rest mass of the test particle becomes a function of the scalar field $\phi$ and hence one has to forgo the principle of equivalence. For a compact review of this and some related issues, we refer to the work of Morganstern\cite{morg}. \\
The intriguing question that is still alive is whether this two descriptions are equivalent or not. Cho\cite{cho1} argued that the Jordan frame is not really the physical description of gravity and the same conclusion holds for Kaluza-Klein theory. Faraoni and Gunzig\cite{faraoni1} arrived at the result that, within the realm of classical framework, Einstein frame is more trustworthy than the Jordan frame. Their work was based upon considerations of gravitational waves. Chiba and Yamaguchi discussed the frame dependence of various cosmological parameters\cite{chiba}. In a more recent work, Faraoni and Nadeau\cite{faraoni2}, however, show that the with some proper interpretation, the two versions are actually equivalent at the classical level. Salgado\cite{salgado} also showed that the apparent mismatch of the Cauchy problem in the two versions can actually be resolved.\\

It is quite expected that the two versions do not give the same physics as they are based upon different physical principles, the principle of equivalence holds in one and does not in the other. Notwithstanding the question of which version is better, practising relativists prefer to work in the Jordan frame so as to be in the comfort zone of the principle of equivalence. For a computational advantage, one might opt for the Einstein frame but while discussing the physical aspects, the metric is transformed back to the Jordan frame via the inverse transformation provided there is no singularity in the solution of the scalar field.\\
Now the bigger question arises regarding the equivalence of the physics obtained in one frame and that in the same frame when the solutions are actually transformed back from the other. The question might appear superfluous, but deserves attention in view of the high degree of nonlinearity in Einstein systems. In the present work, we deal with the question whether these two frames are equivalent even when they are looked at in the same Jordan frame at the quantum level. In the two frames they would look different, but if we transform the solutions in the Einstein frame back to the Jordan frame by effecting the inverse transformation at the level of the solution, would they look the same? We find an answer in the negative! We quantize a spatially flat Friedmann-Robertson-Walker (FRW) cosmological model with a perfect fluid in Brans-Dicke theory in both the versions and pretend that these two are different models altogether. Then we transform the solution for the wave-packet in
the Einstein frame to the Jordan frame via the inverse conformal transformation and check if these two results match. It is quite clearly observed that the results are different. It would have been nice to work only with a BD field, i.e., without a fluid content, but the evolution of the fluid provides a meaningful choice of a properly oriented  time parameter, so the evolution of the system obtained is indeed physically relevant.\\
Recently there has been a similar result, given by Artymowski, Ma and Zhang\cite{ma}, in loop quantum cosmology. It is shown that the quantized version of a spatially flat FRW model in BD theory either in vacuum or in the case of an additional scalar field indeed shows different behaviour in the two frames in the formalism of loop quantum cosmology.\\
In the following section we present the quantum BD model in the Jordan frame and in section 3 the same in Einstein frame. In the last section we discuss the conclusions drawn from the present work.

\section{Brans Dicke theory in Jordan Frame}

The relevant action for Brans-Dicke theory with a perfect fluid can be written as
\begin{equation}
{\cal A} = \int d^4x ~\sqrt{-g} \left[ \phi ~R + \frac{\omega}{\phi} \partial_{\mu} \phi ~\partial^{\mu} \phi \right] + \int d^4x \sqrt{-g} ~P~~,
\end{equation}
where $R$ is the Ricci scalar, and $\phi$ is the BD scalar field which is non-minimally coupled to gravity and $\omega$ is the dimensionless BD parameter. 
Here units are so chosen that $c = 16\pi G_{0} =\hbar=1$. The last term in the equation (1) represents the perfect fluid contribution to the action where $P$ is the pressure and is related to the energy density by the equation of state 
\begin{equation}
P=\alpha \rho ~~,
\end{equation} 
where $\alpha \leq 1$. This restriction on $\alpha$ stems from the consideration that sound waves cannot propagate faster than light. We shall work in a spatially flat spacetime given by the metric
\begin{equation}
\label{FRW}
ds^2 =  n^2(t)dt^2 - a^2(t) dl^{2} ~~,
\end{equation}
where n(t) is called the lapse function, and $dl^{2}$ is the flat 3-space metric. In Schutz's formalism \cite{sch1,sch2}, the fluid four velocity can be expressed in terms of some thermodynamic potentials. Using the normalization of the velocity vector 
\begin{equation}
\label{nor}
u_{\mu}~u^{\mu} = 1,
\end{equation}
one can write the pressure $P$ in any spacetime without rotation as 
\begin{equation}
P=\frac{\alpha}{{(1+\alpha)}^{1+\frac{1}{\alpha}}}h^{1+\frac{1}{\alpha}}e^{-\frac{S}{\alpha}} \, ,
\end{equation}
and the fluid part of the action takes the form (in a comoving system where $u_{\nu}=(n,0,0,0)$)
\begin{equation}
\label{af}
{\cal A}_f = \int dt\left[n^{-\frac{1}{\alpha}}a^3\frac{\alpha}{(1+\alpha)^{1+\frac{1}{\alpha}}}(\dot\epsilon + \theta\dot S)^{1+\frac{1}{\alpha}}e^{-\frac{S}{\alpha}} \right] \, .
\end{equation}
As none of the quantities mentioned depende on space coordinates, the spatial part of the volume integral $\int d^{3}x$ yields a constant $V_{3}$ which is taken to be unity. This does not infringe upon the generality as the constant will not contribute to the subsequaent variation. Here $\epsilon$, $\theta$, $h$ and $S$ are thermodynamic potentials which determine the velocity vector in Schutz formalism. They all satisfy their own evolution equations. An overhead dot represents a differentiation with respect to the coordinate time $t$. The detailed method can be found in the work of Lapchinskii and Rubakov\cite{lapch}. The method has been subsequently used with a high degree of usefulness by many, such as Alvarenga {\it et al}\cite{alvarenga1, alvarenga2}, Vakili\cite{vakili1}, Majumder and Banerjee\cite{barun}, Pal and Banerjee\cite{sridip1, sridip2}. Particularly, for a scalar field, the method had been utilized by Vakili\cite{vakili}, Majumder\cite{barun1} and Almeida {\it
  et al}\cite{almeida}. A very similar approach of expressing the fluid Lagrangian in terms thermodynamic variables has been utilized very recently by B${\ddot o}$hmer, Tamanini and Wright\cite{tamanini1} and Koivisto, Saridakis and Tamanini\cite{tamanini2}. \\

We effect the canonical transformations,
\begin{eqnarray}
T = -p_Se^{-S}p_\epsilon^{-(1 + \alpha)} \quad , \\
\quad p_T = p_\epsilon^{1+\alpha}e^S \quad , \\
 \quad\bar\epsilon = \epsilon - (1 + \alpha)\frac{p_S}{p_\epsilon} \quad , \\
 \quad \bar p_\epsilon = p_\epsilon \quad ,
\end{eqnarray}
along with $p_S=\theta p_\epsilon$. Here $p_\epsilon =\frac{\partial {\cal L}_f}{\partial\dot\epsilon}$, $p_S =\frac{\partial {\cal L}_f}{\partial\dot S}$ and ${\cal L}_f$, the Lagrangian density of the fluid, is the expression inside the square bracket of equation (\ref{af}). The corresponding Hamiltonian for this perfect fluid can be written as
\begin{equation}
{\cal H}_f=n~a^{-3\alpha}~p_T \, .
\end{equation}

The advantage of using this canonically transformed version is that we could find a set of variables where the system of equations is a lot more tractable, while the Hamiltonian structure of the system remains intact. For a discussion regarding the Hamiltonian structure in terms of Poisson brackets, see ref \cite{sridip1}. 

\par
For the spatially flat FRW model given by the metric (\ref{FRW}), the Ricci scalar can be written as
\begin{equation}
R = \frac{1}{a^2~n^3} \left[ -6 a \dot{a} \dot{n} + 6 n \dot{a}^2 + 6 n a \ddot{a} \right] ~~,
\end{equation}
where an overhead dot indicates a differentiation with respect to time $t$. With this $R$, the Lagrangian density  ${\cal L}_g$ for the gravity sector becomes
\begin{equation}
{\cal L}_g = - \frac{6 a \dot{a}^2 \phi}{n} - \frac{6 a^2 \dot{\phi}\dot{a}}{n} + \frac{\omega a^3 \dot{\phi}^2}{n \phi} ~~.
\end{equation}
Using a pair of new variables $q$ and $r$ in place of $a$ and $\phi$, given by,
\begin{center}
 $a=e^{b(q-r)}$ and $\phi=e^{c(q+r)},$ 
\end{center}
where $b$ and $c$ are constants, one can write ${\cal L}_g$ as
\begin{eqnarray}
{\cal L}_g = e^{q(3b+c)}~e^{r(c-3b)} \bigg[ \dot{q}^2 \left( -\frac{6b^2}{n} - \frac{6bc}{n}
+ \frac{\omega c^2}{n} \right) +  \nonumber \\
 \dot{r}^2 \left( -\frac{6b^2}{n} + \frac{6bc}{n} + \frac{\omega c^2}{n} \right) 
 + 2 \dot{q}~\dot{r} \left( \frac{6b^2}{n} + \frac{\omega c^2}{n} \right) \bigg] ~~.
\end{eqnarray}

It deserves mention that this transformation of variables from ($a, \phi$) to ($q, r$) is not a canonical transformation and one has to transform back to the old variables for any physical interpretation. This transformation is effected only to facilitate a separation of variables. \\
In what follows, we shall work with a particular choice $\omega = -\frac{6 b^2}{c^2}$ for the BD parameter $\omega$. There is no significance of this value, this is done only to facilitate the integration so that we can talk about an analytical solution of the Wheeler-DeWitt equation. The Hamiltonian for the gravity sector can be found out from the expression for the Lagrangian ${\cal L}_g$ and the net or super Hamiltonian for the minisuperspace can be written as

\begin{eqnarray}
{\cal H} = {\cal H}_{g} + {\cal H}_{f} = n~ e^{-(3b+c)q}~e^{(3b-c)r} \bigg[ \frac{p_q^2}{24b(-2b-c)} + \nonumber \\
\frac{p_r^2}{24b(-2b+c)} + e^{(3b+c-3\alpha b)q}~e^{(c-3b+3\alpha b)r} p_T \bigg] ~~.
\end{eqnarray}
Here $n$ acts as a Lagrange multiplier taking care of the classical constraint equation ${\cal H}=0$. Using the usual quantization procedure \cite{r6,r7}, we write the Wheeler-DeWitt equation for our super Hamiltonian with the ansatz that the super Hamiltonian operator annihilates the wave function,
\begin{equation}
\label{wdwe1}
\hat{{\cal H}}\quad \vert\Psi(q, r, T)\,\rangle=0.
\end{equation}
We now promote the variables to operators such as $p_{x_i}\rightarrow -i\partial_{x_i}$, $p_T \rightarrow i\partial_T.$  With a particular choice of operator ordering we solve the eqn (\ref{wdwe1}) by the method of separation of variables. It is analytically easy to solve the Wheeler-DeWitt equation with $\alpha=-1$ and $\alpha=0$. 

\begin{itemize}
\item[Case 1 $\rangle$] $\alpha=-1$, $c=6b$ and $\omega = -\frac{1}{6}$
\end{itemize}
\begin{eqnarray}
\Psi (q, r, T) = e^{-iET} \left[ c_1 e^{\frac{kr}{\sqrt{2}}} + c_2 e^{-\frac{kr}{\sqrt{2}}} \right] \nonumber \\
\left[ c_3 J_{\frac{k}{c}}\left( \sqrt{\frac{16E}{3}} e^{cq} \right)  + c_4 J_{-\frac{k}{c}}\left( \sqrt{\frac{16E}{3}} e^{cq} \right) \right]~~. 
\end{eqnarray}
Here $E$ and $k$ are the constants from the separation of variables and $c_i's$ are integration constants with $J$ as the Bessel function of first kind.
If written as a function of $a$, $\phi$ and $T$, we get \\
\begin{eqnarray}
\Psi (a, \phi, T) = e^{-iET} \times \nonumber \\
\left[ c_1 \left( \frac{\phi^{1/6}}{a}\right)^{\frac{k}{2\sqrt{2}b}} + c_2 \left( \frac{a}{\phi^{1/6}}\right)^{\frac{k}{2\sqrt{2}b}} \right]
 \times \nonumber \\
 \left[ c_3 J_{\frac{k}{6b}}\left( \sqrt{\frac{16E}{3}} a^3\sqrt{\phi} \right) + c_4 J_{-\frac{k}{6b}}\left( \sqrt{\frac{16E}{3}} a^3\sqrt{\phi} \right) \right].
\end{eqnarray}

\begin{itemize}
\item[Case 2 $\rangle$] $\alpha=0$, $c=3b$ and $\omega = -\frac{2}{3}$ 
\end{itemize}

\begin{eqnarray}
\Psi (q, r, T) = e^{-iET} \left[ c_1 e^{\frac{kr}{\sqrt{5}}} + c_2 e^{-\frac{kr}{\sqrt{5}}} \right]  \nonumber \\
\left[ c_3 J_{\frac{k}{c}}\left( \sqrt{\frac{40E}{3}} e^{cq} \right) + c_4 J_{-\frac{k}{c}}\left( \sqrt{\frac{40E}{3}} e^{cq} \right) \right]. 
\end{eqnarray}

As a function of $a$, $\phi$ and $T$, this becomes \\

\begin{eqnarray}
\Psi (a, \phi, T)  = e^{-iET} \times \nonumber \\
\left[ c_1 \left( \frac{\phi^{1/3}}{a}\right)^{\frac{k}{2\sqrt{5}b}} + c_2 \left( \frac{a}{\phi^{1/3}}\right)^{\frac{k}{2\sqrt{5}b}} \right]
\times \nonumber \\
\left[ c_3 J_{\frac{k}{3b}}\left( \sqrt{\frac{40E}{3}} a^{\frac{3}{2}}\sqrt{\phi} \right) + c_4 J_{-\frac{k}{3b}}\left( \sqrt{\frac{40E}{3}} a^{\frac{3}{2}}\sqrt{\phi} \right) \right]. 
\label{nordust}
\end{eqnarray}

The solution for the wave-packet can be obtained by a linear superposition of the eigenfunctions. We do that for a dust distribution ($\alpha =0$) for the second case ($c=3b$, $\omega = -\frac{2}{3}$) and get, 
\begin{eqnarray}
\Psi_{wp}(a,\phi,T) = \int_{r'=0}^{\infty} \int_{k=0}^{1} f(r',k) ~r'~ e^{-\frac{i 3r'^2 T}{40}} \times \nonumber \\
\left(\frac{a}{\phi^{\frac{1}{3}}}\right)^{\frac{k}{2\sqrt{5}b}} J_{\frac{k}{c}}(r' a^{\frac{3}{2}}\sqrt{\phi}) ~dr' ~dk ~~.
\label{wp1}
\end{eqnarray}

Here we have considered $c_1=c_4=0$ (from eqn (\ref{nordust})) to satisfy the boundary condition imposed on $\Psi \vert_{a=0}=0$. A change in variable $r'=\sqrt{\frac{40E}{3}}$ is also considered. The function $f(r',k)$ is a suitable weight function for the construction of the wave packet. If we consider $f(r',k)=e^{-\alpha^{\prime}r'^2}r'^{\frac{k}{c}}$, where $\alpha^{\prime}$ is a positive quantity. The integrals of (\ref{wp1}) can be analytically evaluated to yield \cite{bab13}.

\begin{equation}
\Psi_{wp}(a,\phi,T) = \frac{e^{-\frac{a^3 \phi}{4\alpha_g}}~ \left\{1 - (2\alpha_g)^{\frac{-1}{c}}~ a^{\frac{1+\sqrt{5}}{2\sqrt{5}b}} ~ \phi^{\frac{\sqrt{5}-1}{6\sqrt{5}b}}\right\}}  {\alpha_g ~ \ln( 4 \alpha_g^2 ~ a^{-\frac{3+3\sqrt{5}}{\sqrt{5}}} ~ \phi^{\frac{1-\sqrt{5}}{\sqrt{5}}})} ~~.
\end{equation}

\begin{figure*}[t]
\begin{center}
\includegraphics[width=8.0cm,clip=true]{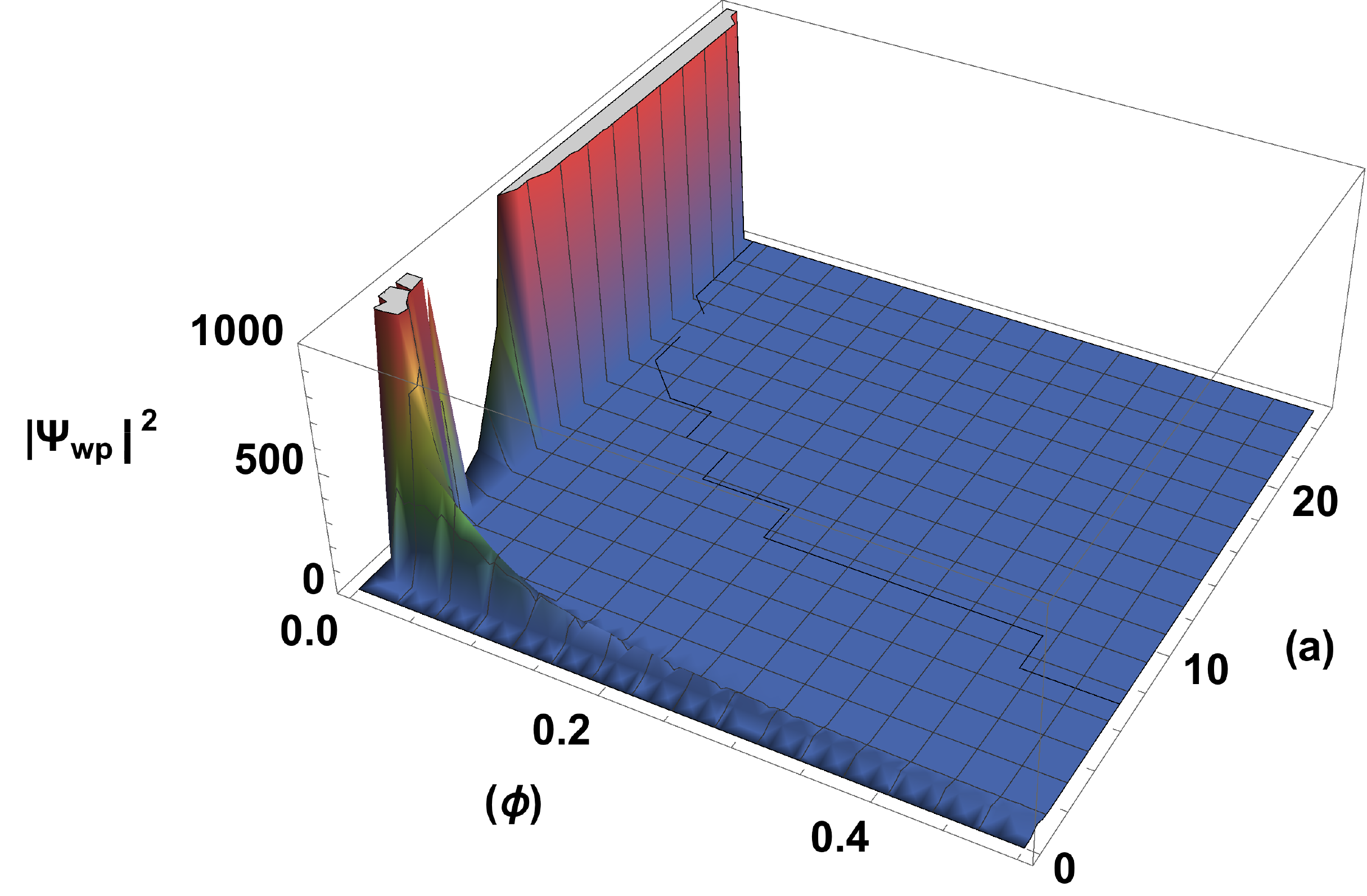} 
\hspace{0.5cm}
\includegraphics[width=8.0cm,clip=true]{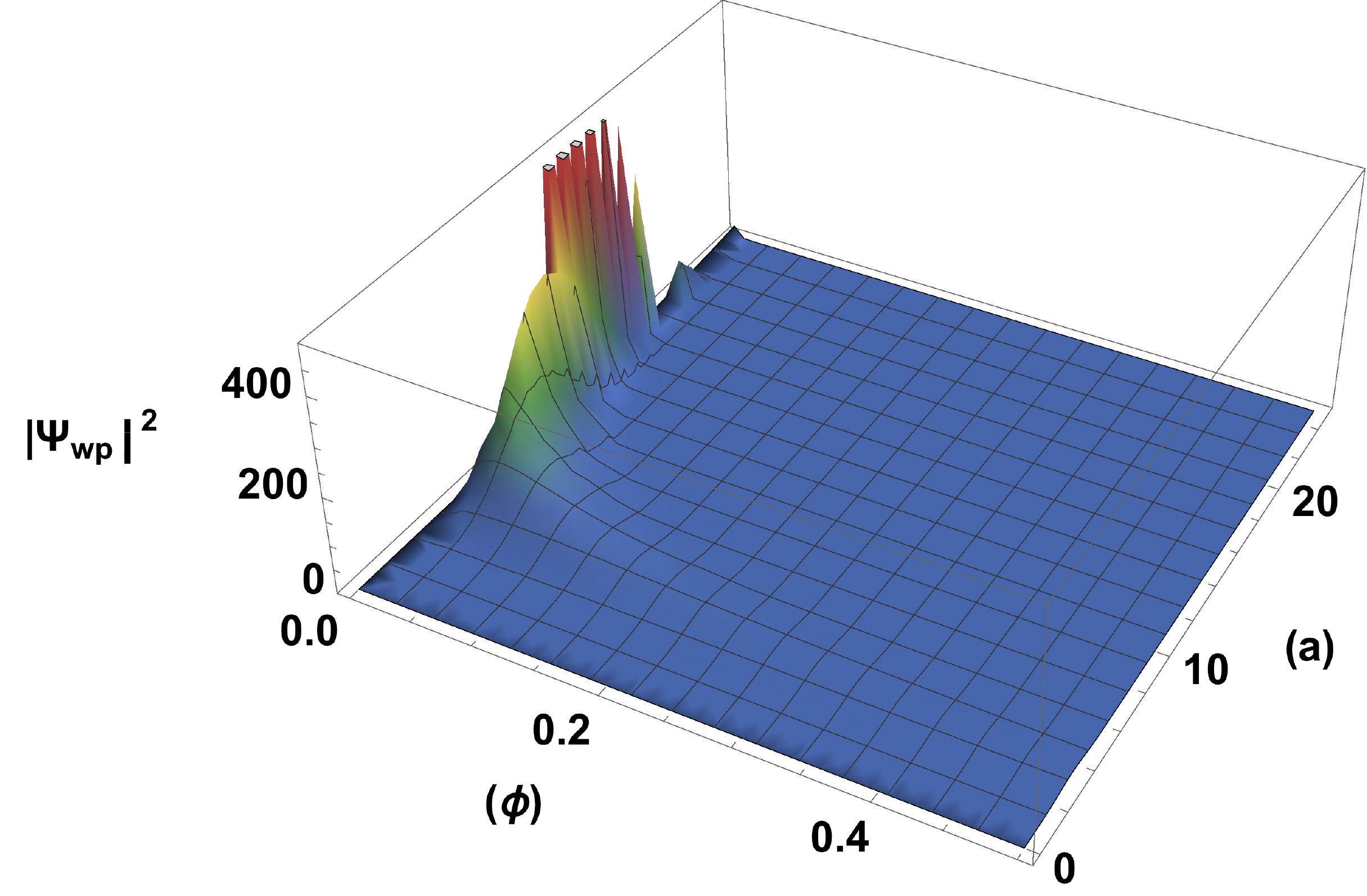}  
\caption{\label{modpsi} (Color Online) Here we have plotted the nature of $\vert \Psi_{wp} \vert^2$ of Eqn. (\ref{modpsisq1}) as a function of $a$ and $\phi$ at different values of $T$. The {\it left} figure is at $T=0$ and the {\it right} is at $T=10$ in some arbitrary units. We have considered $\alpha^{\prime}=0.1$ and $b=\frac{1}{\sqrt{5}}$ with $c=3b$. This is in Jordan frame.}
\end{center}
\end{figure*}

First we integrate with respect to $r'$ using the known integral for the Bessel function $\int_0^{\infty} e^{-mx^2}x^{n+1}J_n(px)dx=\frac{p^n}{(2m)^{n+1}}e^{-\frac{p^2}{4m}}$ such that the second integral becomes a known gaussian integral. Here $\alpha_g=\alpha^{\prime}+\frac{i3T}{40}$. An overall constant accompanying $\Psi_{wp}$, coming from several steps in the calculation, is taken to be unity without any loss of generality. We can also calculate $\vert \Psi_{wp} \vert^2$, the norm of the wave packet as

\begin{equation}
\label{modpsisq1}
\vert \Psi_{wp} \vert^2 = \frac{A_3^2}{A_1^2 ~A_2^2}~ e^{-\frac{8 \alpha^{\prime} a^3 \phi}{16\alpha^{\prime 2}+\frac{9T^2}{100}}} ~~,
\end{equation}

where 
\begin{equation}
A_1 = \sqrt{\alpha^{\prime 2}+\frac{9T^2}{1600}} ~~, \nonumber 
\end{equation}
\begin{equation}
A_2 = \left[\left(\ln \frac{4 A_1^2~ \phi^{\frac{1}{\sqrt{5}}-1}}{a^{\frac{3}{\sqrt{5}}+3}}\right)^2 ~+ 4 \left(\tan^{-1} \frac{3T}{40\alpha^{\prime}}\right)^2\right]^{\frac{1}{2}} ~~, \nonumber
\end{equation}
and
\begin{eqnarray}
A_3 = \bigg[1 - 2 (2A_1)^{\frac{-1}{c}} ~ a^{\frac{1+\sqrt{5}}{2\sqrt{5}b}} ~ \phi^{\frac{\sqrt{5}-1}{6\sqrt{5}b}}~ \cos\left(\frac{\tan^{-1} \frac{3T}{40\alpha^{\prime}}}{c}\right) \nonumber \\
+ (2A_1)^{-\frac{2}{c}} ~a^{\frac{\sqrt{5}+1}{\sqrt{5}b}} ~ \phi^{\frac{\sqrt{5}-1}{3\sqrt{5}b}} \bigg]^{\frac{1}{2}} ~~. 
\end{eqnarray}
In Fig.(\ref{modpsi}) we plot the nature of $\vert \Psi_{wp} \vert^2$ as a function of $a$ and $\phi$.

\section{Brans-Dicke theory in Einstein Frame}

\begin{figure*}[t]
\begin{center}
\includegraphics[width=8.0cm,clip=true]{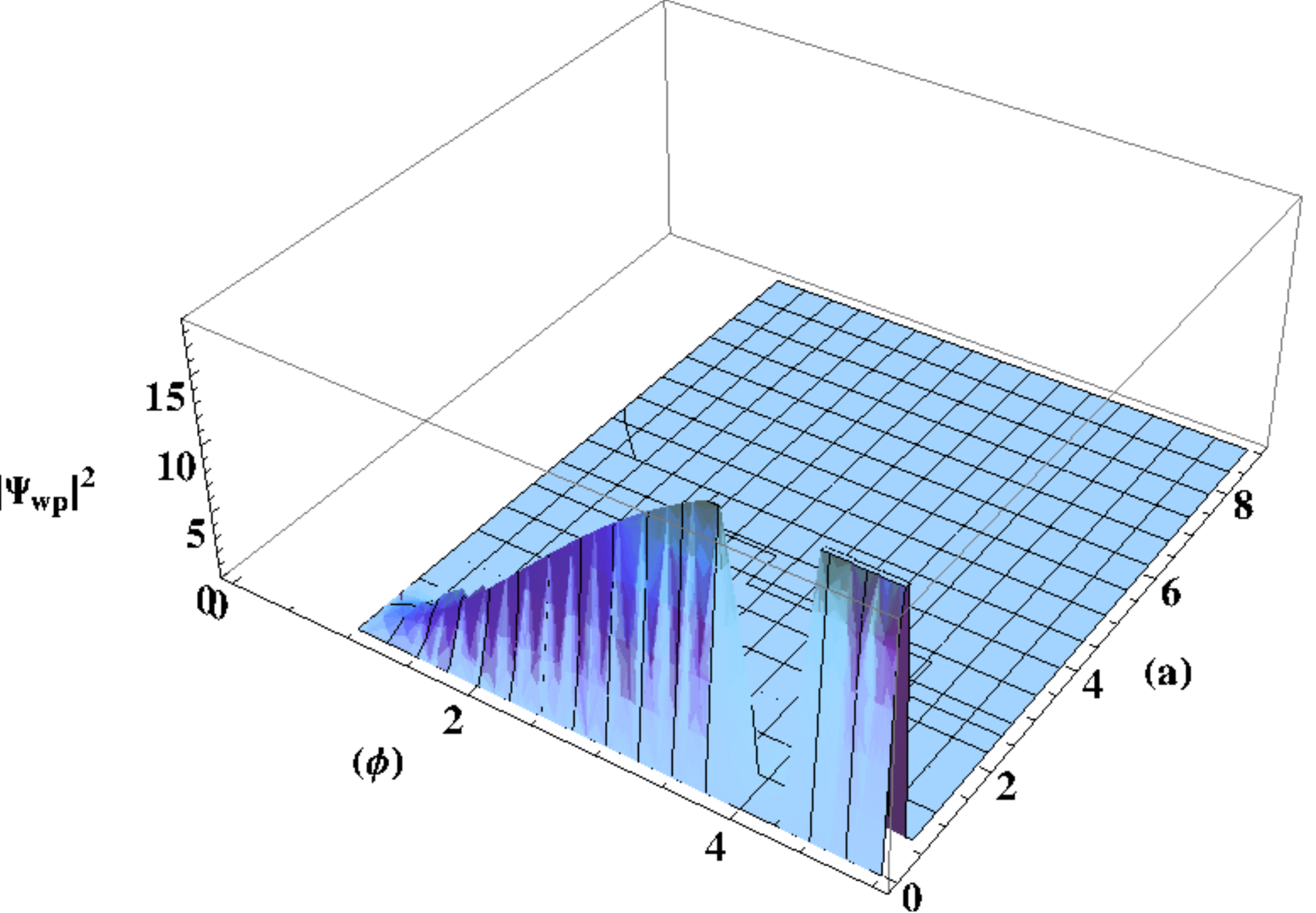} 
\hspace{0.5cm}
\includegraphics[width=8.0cm,clip=true]{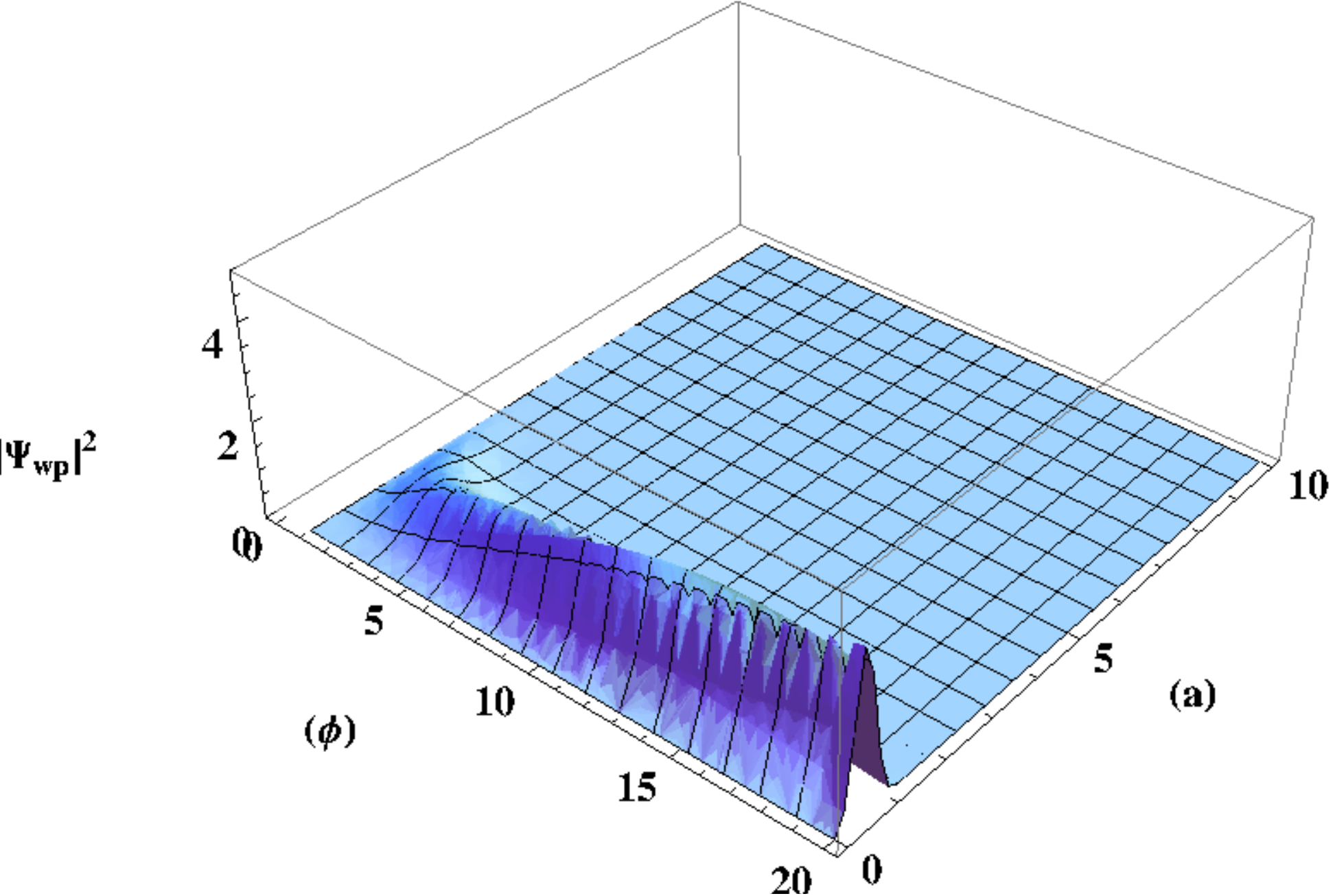}  
\caption{\label{modpsi2} (Color Online) Here we have plotted the nature of $\vert \Psi_{wp} \vert^2$ of Eqn. (\ref{modpsisq2}) as a function of $a$ and $\phi$ at different values of $T$. The {\it left} figure is at $T=0$ and the {\it right} is at $T=10$ in some arbitrary units. We have considered $\alpha^{\prime}=0.1$. This is in Einstein frame.} 
\end{center}
\end{figure*}

The relevant action for the theory with a perfect fluid can be written as
\begin{eqnarray}
{\cal A} = \int d^4 x ~\sqrt{-\bar{g}} \left[ \bar{R} + \left(\omega + \frac{3}{2} \right) {\bar g}^{\mu \nu} \partial_{\mu} \xi ~\partial_{\nu} \xi \right] \nonumber \\
+ \int d^4x \sqrt{-\bar{g}} ~{\bar P}~~,
\end{eqnarray}
where $\bar{R}$ is the Ricci scalar, and $\xi$ is a massless scalar field which is minimally coupled to gravity in the revised version of the theory. One can arrive at this action with the aid of a conformal transformation $\bar{g}_{\mu\nu}=\phi g_{\mu\nu}.$ The scalar field $\xi = \ln \phi.$ The quantities appearing in this action are in the transformed version, an overhead bar indicates that.\\
In a flat FRW background it is rather easy to calculate the super Hamiltonian constraint and it can be written as
\begin{equation}
{\cal H} = -\frac{n~p_{\bar{a}}^2}{24\bar{a}} + \frac{1}{4} \frac{n \xi^2 ~p_{\xi}^2}{\left(\omega + \frac{3}{2}\right)\bar{a}^3} + \frac{n~p_{\bar{T}T}}{\bar{a}^{3\alpha}} ~~.
\end{equation} 
The coefficient of $p_{\xi}^{2}$ contains an additional ${\xi}^{2}$ as we shall work with the coordinates $x^{\mu}$ of the Jordan frame so that the final comparison is consistent. This Hamiltonian is consistent with that given by Zonghong\cite{zong} (see also \cite{zong1}). For the relevant transformations, we refer to \cite{dicke}. Here also the Schutz formalism has been utilized. Following the same procedure as before, we solve the Wheeler-De Witt equation with $\alpha=-1$ and $\alpha=0$. The stationary wave functions in the revised version are given as:

\begin{itemize}
\item[Case 1 $\rangle$] $\alpha=-1$ and $\omega = -\frac{1}{6}$ 
\end{itemize}
\begin{eqnarray}
\Psi (\bar{a}, \xi, T)  = e^{-iET} \times \nonumber \\
\sqrt{\bar{a} \xi} \left[ c_1 \xi^{\sqrt{\frac{1}{4}+\frac{k^2}{6}}} + c_2 \xi^{-\sqrt{\frac{1}{4}+\frac{k^2}{6}}} \right]\times \nonumber \\
\left[ c_3 J_{\frac{\sqrt{1+3k^2}}{6}}\left( \sqrt{\frac{8E}{3}} \bar{a}^3 \right) +  c_4 J_{\frac{-\sqrt{1+3k^2}}{6}}\left( \sqrt{\frac{8E}{3}} \bar{a}^3 \right) \right]
\end{eqnarray}
\begin{itemize}
\item[Case 2 $\rangle$] $\alpha=0$ and $\omega = -\frac{2}{3}$ 
\end{itemize}
\begin{eqnarray}
\Psi (\bar{a}, \xi, T)  = e^{-iET} \times \nonumber \\
\sqrt{\bar{a} \xi} \left[ c_1 \xi^{\sqrt{\frac{1}{4}+\frac{k^2}{6}}} + c_2 \xi^{-\sqrt{\frac{1}{4}+\frac{k^2}{6}}} \right] \times \nonumber \\
\left[ c_3 J_{\frac{\sqrt{5+24k^2}}{3\sqrt{5}}}\left( \sqrt{\frac{32E}{3}} \bar{a}^{\frac{3}{2}} \right) + c_4 J_{-\frac{\sqrt{5+24k^2}}{3\sqrt{5}}}\left( \sqrt{\frac{32E}{3}} \bar{a}^{\frac{3}{2}} \right) \right]
\end{eqnarray}

We have used some of the notations ($\Psi, E, k$) same as that of the last section, but they are not same and only related to the respective differential equation. We take up the $\alpha=0$ case in this transformed version as well. We will consider $c_2=c_4=0$ so as to have the similar boundary conditions like the example in the Jordan frame and get 
\begin{eqnarray}
\Psi (a,\phi,T)  \propto e^{-iET} ~ \phi^{\frac{1}{4}} ~ (a \ln \phi)^{\frac{1}{2}}~ (\ln \phi)^{\sqrt{\frac{1}{4}+\frac{k^2}{6}}}~ \times \nonumber \\
J_{\frac{\sqrt{5+24k^2}}{3\sqrt{5}}} \left(\sqrt{\frac{32E}{3}}a^{\frac{3}{2}}\phi^{\frac{3}{4}}\right) ~~,
\end{eqnarray}
as $\bar{a}=a\sqrt{\phi}$ and $\xi=\ln \phi$. The wave-packet can be constructed by the superposition of the eigenfunctions as  
\begin{eqnarray}
\Psi_{wp} = \sqrt{\bar{a}\xi} \int_{r'=0}^{\infty}\int_{s=0}^{1} e^{-\alpha_g r^2}~ r^{s+1}~ J_s(r\bar{a}^{\frac{3}{2}})~\xi^{\sqrt{\frac{31}{144}+\frac{5s^2}{16}}}~
\nonumber \\
\times \frac{s}{\sqrt{\frac{15s^2}{8}-\frac{5}{24}}} ~ds ~dr' ~~,
\end{eqnarray}
where $\alpha_g = \alpha^{\prime} + \frac{i3T}{32}$, $s=\frac{\sqrt{5+24k^2}}{3\sqrt{5}}$ and we have incorporated a quasi-Gaussian weight factor. If we make the approximation $\sqrt{\frac{15s^2}{8}-\frac{5}{24}} \sim \sqrt{\frac{15}{8}}s$ and $\sqrt{\frac{31}{144}+\frac{5s^2}{16}} \sim \frac{\sqrt{5}}{4}s$ then the above integral can be evaluated in a closed form \cite{bab13}. The approximations are only meant to evaluate the integrals analytically such that we can compare with results found in the last section. Upto a constant proportionality factor, which comes from several steps of integration, we get
\begin{equation}
\Psi_{wp} = \frac{\phi^{\frac{1}{4}}~ \sqrt{a\ln \phi} ~e^{-\frac{a^3 \phi^{\frac{3}{2}}}{4\alpha_g}}
\{2\alpha_g - a^{\frac{3}{2}}~\phi^{\frac{3}{4}}~(\ln \phi)^{\frac{\sqrt{5}}{4}}\}}
{\alpha_g^2 \ln \left(\frac{16\alpha_g^4}{a^6 \phi^3 (\ln \phi)^{\sqrt{5}}}\right)} ~~,
\end{equation}
where, using the inverse transformation, $\bar{a}$ and $\xi$ are replaced by $a$ and $\phi$ respectively. We can also calculate $\vert {\Psi}_{wp} \vert^2$ aimed for finding the norm of the wave packet,
\begin{equation}
\label{modpsisq2}
\vert {\Psi}_{wp} \vert^2 = \frac{B_2^2 a \sqrt{\phi} \ln \phi}{B_1^4 B_3^2} ~ e^{-\frac{8 \alpha^{\prime} a^3 \phi^{\frac{3}{2}}}{16\alpha^{\prime 2}+\frac{9T^2}{64}}}
\end{equation}
where
\begin{equation}
B_1 = \sqrt{\alpha^{\prime 2} +\frac{9T^2}{1024}} ~~, \nonumber
\end{equation}
\begin{equation}
B_2 = \left[ \{2\alpha^{\prime} - a^{\frac{3}{2}} \phi^{\frac{3}{4}} (\ln \phi)^{\frac{\sqrt{5}}{4}}\}^2 + \frac{9T^2}{256}\right]^{\frac{1}{2}} ~~, \nonumber
\end{equation}
and
\begin{equation}
B_3 = \left[ \left\{ \ln \frac{16 B_1^4}{a^6 \phi^3 (\ln \phi)^{\sqrt{5}}}\right\}^2 + 16 \{\tan^{-1} \frac{3T}{32\alpha^{\prime}}\}^2\right]^{\frac{1}{2}} ~~.
\end{equation}
In Fig.(\ref{modpsi2}) we plot the nature of $\vert \Psi_{wp} \vert^2$ as a function of $a$ and $\phi$.

\section{Discussion}

If we compare equations (20) and (31), the solution for the Wheeler DeWitt equations for the dust distribution ($\alpha =0$) for a particular value of the BD parameter $\omega$ as $-\frac{2}{3}$, we find that the solutions are evidently different. One should emphasize that this comparison is made after the Einstein frame solution is transformed back to the Jordan frame via the inverse transformation, $g_{\mu\nu}={\phi}^{-1}{\bar g}_{\mu\nu}$. The solutions are intricate, so it is perhaps better to look at some physical aspects of the solutions. The norm of the wave packets are found in the two versions. They are given by equations (23) and (34) for the one that calculated in the Jordan frame directly and the one that calculated in the Jordan frame from the solutions transformed back from the Einstein frame respectively. The norms are depicted in figures (\ref{modpsi}) and (\ref{modpsi2}) respectively for two epochs of time. It is once again quite evident from the figures that
  the corresponding norms are qualitatively different, peaked at different locations.
\par
Thus it is conclusively established that the two frames are physically different at least at the quantum level. It is true that this work shows this in one example. But one counter example is good enough to show the non-equivalence of the two frames. The particular values of $\alpha$ and $\omega$ chosen are indeed for the sake of analytical calculations, but the values are quite legitimate, $\alpha =0$ represents a pressureless dust, and $\omega = -\frac{2}{3}$ is neither pathological nor trivial. In fact, as evident from the action functional (27), $\omega = - \frac{3}{2}$ is the trivial case as it indicates a zero kinetic energy for the BD field and values of $\omega$ less than that would indicate a pathology of a negative kinetic energy.
\par
We have solved the Wheeler DeWitt equation for another case as well, namely $\alpha = -1$ indicating an effective cosmological constant with another legitimate value of $\omega = -\frac{1}{6}.$ But the comparison of the two frames were not possible for the difficulty in the integration. However, if we compare the solution of the Wheeler-DeWitt equations, namely the solution given in equation (18) with that in (29), it is rather apparent that they would not match after effecting the inverse transformation in the latter. This indeed lends a support towards the claim of the present work regarding the nonequivalence of the two frames. 
\par
In agreement with the recent result in loop quantum cosmology\cite{ma} , the present work, in a different philosophy of quantization, explicitly shows that Jordan frame and Einstein frames are physically different. Our result is also consistent with the finding of Faraoni and Nadeau\cite{faraoni2} that the two frames are not equivalent at the quantum level.
\par
Certainly there is ample scope of improvement on the present work. An important issue is that there is nothing to check that the Hamiltonian written in the two frames correspond to each other so far as the operator ordering is concerned. The only thing that could be taken care of in terms of correspondence is the fact that in both cases the functions of the coordinates come first followed by the momenta. We also took care to fix the boundary conditions in some way. But question remains if that is enough. \\
\par
But the question of this equivalence is important, and until this is settled, this kind of indications have to be relied upon in the absence of better results. \\

{\bf Acknowledgement:} The authors are grateful to ananimous referee for many important and thought provoking comments which helped us in improving the manuscript by a great deal. \\

\end{document}